\documentstyle[aps]{revtex}
\begin{document}

\title{Matrix string interactions}
\author{Vipul Periwal and \O yvind Tafjord}
\address{Department of Physics,
Princeton University,
Princeton, New Jersey 08544}

\def\dd{\hbox{d}}
\def\tr{\hbox{tr}}\def\Tr{\hbox{Tr}}
\def\ee#1{{\rm e}^{{#1}}}
\maketitle
\begin{abstract}
String configurations have been identified in compactified Matrix
theory at vanishing string coupling.
We show qualitatively
how the interactions of these strings are determined by the
Yang-Mills gauge field on the worldsheet. At finite string coupling,
this suggests the underlying dynamics is not well-approximated as a
theory of strings. This may explain why string perturbation theory
diverges badly, while Matrix string perturbation theory presumably
has a perturbative expansion with properties similar to the strong
coupling expansion of 2d Yang-Mills theory. \end{abstract}
String perturbation theory diverges. This divergence has been linked
by Shenker\cite{shenker}
to the unusual strength of non-perturbative effects in string theory.
Witten\cite{witten} has pointed out that the existence of
Ramond-Ramond charged states may lead to precisely such
non-perturbative effects.

Following the conjecture of
Banks, Fischler, Shenker and Susskind\cite{bfss}, and using the
result of Taylor\cite{wati}, string configurations have been
identified in 2d maximally supersymmetric Yang-Mills
theory\cite{motl,banks,dvv}. For another approach to string
configurations in Yang-Mills theory, see \cite{ikkt}. What is
remarkable about these models\cite{motl,banks,dvv} is that weak
string coupling is related to strong Yang-Mills coupling, most
explicitly stated in \cite{dvv}.
Strong-coupling expansions typically have a finite radius of
convergence\cite{dz}, so it
is of interest to identify how string perturbation theory differs
from Matrix string perturbation theory.
To address this issue, it is necessary to identify how the
interactions of Matrix strings arise from the underlying Yang-Mills
theory.

For concreteness, we
focus on the identification of string configurations in the limit of
vanishing string coupling proposed by Dijkgraaf, Verlinde and
Verlinde\cite{dvv}. An equivalent description of these configurations
was given in \cite{motl,banks}. Recall that, according to
Taylor\cite{wati}, Matrix theory\cite{bfss} compactified on a circle
is described by a 2d sYM theory on ${\bf R}\times $ the dual circle.
The action is just the dimensionally reduced action obtained from 10d
sYM theory. The 8 components of the 10d gauge potential become scalar
matter fields on the 2d worldsheet, corresponding to the the
transverse space coordinates of the string in lightcone gauge. The
action of interest is
\[ S= {1\over 2\pi\alpha'}
\int \Tr\left(D_{\mu}XD^{\mu}X + \Theta^{T} \Gamma^{\mu}D_{\mu}\Theta
+ g_{s}^{2}F_{\mu\nu}F^{\mu\nu} + {1\over g_{s}^{2}}
[X^{i},X^{j}]^{2} + {1\over g_{s}}
\Theta^{T}\Gamma^{i}[X_{i},\Theta]\right),\] where $i=1,\ldots,8,$
$\mu=0,1,$ are worldsheet indices, and $\Theta$ is a Majorana-Weyl
spinor in 10d. The worldsheet is taken to be cylindrical with
$\sigma\in [0,2\pi].$ This is to be considered for large $N,$ where
U($N$) is the gauge group, following \cite{bfss}.

It was argued in \cite{motl,banks,dvv} that at $g_{s}=0$ the
equations of motion imply that the $X$ matrices mutually commute,
with $[\Theta, X]=0,$ and hence may be diagonalized simultaneously.
The limit $g_{s}=0$
should correspond to an infrared fixed point, and hence to a
superconformal field theory. As one goes from $\sigma=0$ to
$\sigma=2\pi,$ the eigenvalues may have nontrivial monodromy, in the
sense that the eigenvalues only label orbits of U($N$) up to
permutations, so if $\lambda^{i}$ stands for the diagonalized $X^{i}$
matrices, \[ \lambda(\sigma=0) = P\lambda(\sigma=2\pi)P^{{-1}}, \]
for some permutation matrix $P$ in the defining representation of
$S_{N}.$
Ref.~\cite{motl,banks,dvv} suggested that string configurations
should correspond to
cycles in this permutation.

The interaction of strings in the model was explained in \cite{dvv}
as arising from the restoration of a non-Abelian U(2) subgroup when
two eigenvalues coincide. Surprisingly, the description of this
interaction given in \cite{dvv}, in
terms of a twist field in the superconformal field theory that
describes the IR fixed point at $g_{s}=0,$ makes {\it no} reference
to the Yang-Mills Lagrangian. Of course, if there is no contact with
the Yang-Mills Lagrangian, the Matrix approach would be a
particularly obtuse way of thinking about light-cone string
perturbation theory (which is not an enlightening approach to string
theory in the first instance). In particular, it is unclear why
configurations of a few long
strings should dominate the dynamics from the given description of
string interactions. The only non-perturbative content in the model
is the Yang-Mills action, and one would like to {\it derive} the
conjectured string interaction from this action. This should also
dynamically
determine which configurations of strings actually dominate the
dynamics.

We show in this note how the Yang-Mills gauge field on the worldsheet
determines the dominant transitions between different string
configurations in this model. In doing so, we find that the
eigenvalue description is really only suited to $g_{s}=0.$ Away from
$g_{s}=0,$ the physics is much clearer in terms of the full matrices.
We believe this is the reason why string perturbation theory diverges
badly---it is simply that the description in terms of just string
configurations, {\it i.e.} the eigenvalues in \cite{motl,banks,dvv},
is valid
only at $g_{s}=0$!

We are interested in the sYM theory at strong Yang-Mills coupling.
Expanding $S$ about an ultralocal theory, we observe that, due to
supersymmetric cancellations, the first quantum corrections to the
action appear at order $g_{s}^{2} $ for configurations with commuting
$X$ matrices.  It is therefore meaningful to consider the effects of
classical terms of order $g_{s}^{0}.$ These terms imply that if the
$X$ matrices commute, $A$ is that connection such that the $X$
matrices are covariantly constant. Since the $X$ matrices are
sections of a twisted bundle, with monodromy $P,$ the holonomy of the
gauge connection must also be $P$ for minimizing the
$D_{\mu}XD^{\mu}X$ term in the action. While we expressed this in the
continuum formulation, it is even simpler to see this from the form
of the lattice covariant derivative, $D_{\mu}X(x) \equiv
U_{x,\mu}X(x+e_{\mu})U_{x,\mu}^{-1} - X(x).$

Since the 2d sYM theory
is strongly-coupled, an appropriate starting point for calculations
is the lattice theory.  Continuum
weak-coupling engineering dimensions of operators are not meaningful
in such a lattice theory, so we will not be able to derive the
dimension of the operator in the conformal theory ($g_{s}=0$) that
generates interactions. It is possible, however, to ask qualitative
questions about the nature of interactions implied by the Yang-Mills
action.

We deduce the {\it approximate} form of the dominant transitions from
the remaining non-vanishing term in $S$: The transition from a
configuration described by a monodromy $P$ to one with monodromy $P'$
is weighted approximately (in Euclidean lattice gauge theory
terminology) by
\[ \exp\left({g_{s}^{2}\over 2\pi\alpha'} \Tr
(P'P^{-1}+PP'^{-1}-2)\right) \]
in $A^{0}=0$ `gauge'. Thus the curvature of the gauge field induced
by a transition from one monodromy to another is the determining
factor in the relative strengths of different interactions. When the
monodromy matrices are permutation matrices, it is easy to see that
the least possible curvature is induced by a transition in which two
cycles coalesce to form a longer cycle, or vice versa. Comparing this
to the interactions described in \cite{dvv}, we see that the sYM
action provides a qualitative justification for the assumed form of
string interactions.

This is, of course, a gross oversimplification, and should not be
taken to mean that $F^{2}$ is the continuum operator that generates
interactions in the strong-coupling conformal field theory. Indeed,
the spectrum of scaling operators at $g_{s}=0$ has little to do with
the different terms in $S.$ We have only deduced the qualitative form
of the interactions of string configurations from the sYM action. The
other terms that appear at the same order ($g_{s}^{2}$) are (1)
quantum
corrections to the ultralocal theory, and (2) quantum corrections
from fluctuations of the terms that determined the holonomy of the
connection.
To determine the precise operator form of interactions at
$g_{s}\not=0,$ one must include these quantum effects. In
particular, the transition from one string configuration to another
is affected by off-diagonal matrix elements of $X$, and by the
fluctuations of the gauge field on the world-sheet. The dimension 3
string splitting operator\cite{dvv} that
appears exactly at $g_{s}=0$ is a result of all these dynamical
effects.
For $g_{s}>0$, one is dealing with the full matrix structure of the
theory, and the stringy nature is no longer obvious.

For a more systematic exploration of the strong-coupling limit of the
lattice gauge theory, it would be appropriate to rescale the fields
in the action. Written as a dimensional reduction of 10d sYM theory,
we would expect just to see $g_{s}^{2}$ in front of the entire
action. By rescaling
$X,\Theta$ one can write the action as $S$ above, or as \[ S'\equiv
{1\over 2\pi\alpha'}
\int \Tr\left(g_{s}D_{\mu}XD^{\mu}X + \sqrt g_{s} \Theta^{T}
\Gamma^{\mu}D_{\mu}\Theta
+ g_{s}^{2}F_{\mu\nu}F^{\mu\nu} +
[X^{i},X^{j}]^{2} +
\Theta^{T}\Gamma^{i}[X_{i},\Theta]\right),\] a form suitable for an
expansion in $g_{s}$  since all terms involving
derivatives have positive powers of $g_{s}.$   This is equivalent to a
strong-coupling expansion in the Yang-Mills coupling.

The lack of a clear stringy interpretation for Matrix string
interactions
is a good thing, since if the stringy description were to be valid at
$g_{s}\not=0,$ we would be left with firm evidence {\it against}
Matrix theory, given the different perturbative behaviours of strong
coupling Yang-Mills theory, describing Matrix strings, and
garden-variety string theory.
If one could surmount the problem of formulating supersymmetric
lattice gauge theories\cite{golter}, one would have a concrete tool
to understand Matrix string
theory, and see how its behaviour differs from string theory.

Conversations with Wati Taylor and Larus Thorlacius are gratefully
acknowledged.
This work was supported in part by NSF grant PHY96-00258.

\end{document}